\documentclass[preprint,preprintnumbers,amsmath,amssymb,nofootinbib]{revtex4}

\usepackage{amssymb,amsthm,amscd,amsbsy,array}
\usepackage{bm}
\usepackage{soul} 
\usepackage{graphics,graphicx,xcolor}
\usepackage[colorlinks=true, pdfstartview=FitV, linkcolor=blue, citecolor=blue, urlcolor=blue]{hyperref}

\usepackage{amsfonts}
\usepackage{latexsym}
\usepackage{etex}
\usepackage{dcolumn}
\usepackage{amsmath,dsfont}
\usepackage{soul} 

\def\ben{\begin{equation}}
\def\een{\end{equation}}
\def\bea{\begin{eqnarray}}
\def\eea{\end{eqnarray}}

\def\beq{\begin{equation}}
\def\eeq{\end{equation}}
\def\beqa{\begin{eqnarray}}
\def\eeqa{\end{eqnarray}}
\def\nn{\nonumber}
\def\barray{\left(\begin{array}}
\def\earray{\end{array}\right)}
\def\barraynb{\begin{array}}
\def\earraynb{\end{array}}

\def\?{\quad{\gb{\fbox{\texttt{?}}\;}}\quad}

\def\v0{\mathbf{0}}

\usepackage{color}

\newcommand{\gb}{\colorbox{green}}

\def\beq{\begin{equation}}
\def\eeq{\end{equation}}
\def\bea{\begin{eqnarray}}
\def\eea{\end{eqnarray}}

\def\semidirectproduct{
{\ooalign
{\hfil\raise.07ex\hbox{s}\hfil\crcr\mathhexbox20D}}} 

\def\6{\partial}
\def\7{\tilde}
\def\8{\widehat}

\def\pa{\partial}

\def\G11{\Gamma_{11} }

\usepackage{color}

\def\beq{\begin{equation}}
\def\eeq{\end{equation}}
\def\beqa{\begin{eqnarray}}
\def\eeqa{\end{eqnarray}}
\def\nn{\nonumber}

\let\ssection=\section
\renewcommand{\section}{\setcounter{equation}{0}\ssection}

\begin{document}


\title{
L\'evy-Leblond fermions on  the wormhole
\\[6pt]
}

\author{M. Cariglia$^{1}$\footnote
{e-mail:Marco.Cariglia@ufop.edu.br}
G. W. Gibbons$^{2}$\footnote{
mailto:G.W.Gibbons@damtp.cam.ac.uk},
}
\affiliation{
$^{1}${\small DEFIS, Universidade Federal de Ouro Preto,  MG-Brasil}
\\ 
$^2${\small D.A.M.T.P., Cambridge University, U.K.}
\\
}

\date{\today}

\begin{abstract}
We propose a simple model of entanglement generated by geometry, studying non-relativistic massive L\'evy-Leblond fermions in the $1+2$ geometry of a Bronnikov-Ellis wormhole. The model is equivalent to that of relativistic massless Dirac fermions in $1+3$ dimensions, where one spatial direction is flat. The effect of the wormhole is to generate quantum states that, far from the throat, are approximated by entangled particles on two flat, separated spacetime regions. An appealing feature of the model is that it has a condensed matter analogue, the  regime of intermediate energies for two planes of bilayer graphene linked by a bilayer carbon nanotube. Therefore we expect that it might be possible to realize in the laboratory the entangled states studied here. We argue that generalisations of our solvable model which preserve the topology will have similar quantum behaviour. 
\\[40pt]
%
%
\\
\end{abstract}

\maketitle

\baselineskip=16pt

%
%

\section{Introduction} 
The Einstein-Rosen\cite{Einstein:1935tc} =
 Einstein-Poldolksy-Rosen \cite{Einstein:1935rr}  conjecture
\cite{Maldacena:2013xja,Maldacena1,Maldacena:2017axo,Maldacena:2018lmt,Susskind2016},
has generated a lot of interest recently as it links geometrical features of quantum gravity such as wormholes to fundamental properties of quantum mechanical states as entanglement. 
 
Central to the proposal is the idea that nontrivial
topological properties of spacetime backgrounds naturally admit 
entangled quantum states of  quantum fields propagating on
these backgrounds and moreover that these states could in
principle arise dynamically. Most attention has focussed
on black hole spacetimes with or without a negative
cosmological constant which admit a (possibly partial)
Cauchy surface with two asymptotically  flat or asymptotically hyperbolic
ends as first discussed in \cite{Einstein:1935tc} and often referred to as an
Einstein-Rosen Bridge or a wormhole, although
the latter term has been used more generally in the  related but
different sense  of a bridge between two regions of a Cauchy surface with a single end. In fact as early as 1916
Flamm \cite{Flamm}  had  drawn attention to the outer half of the bridge
\cite{Gibbons}. In 1976  Israel \cite{Israel:1976ur}  drew attention to the relevance of considering both sides of the Einstein-Rosen Bridge when considering thermal
Hartle-Hawking type states and in 1986 T.D. Lee
\cite{Lee:1985rp} drew attention to the
relevance of EPR type entanglement.

In this paper we  propose a simple model where entangled states of
non-relativistic fermions are generated geometrically via a
Bronnikov-Ellis \textcolor{red}{\cite{Bronnikov:1973fh,Ellis:1973yv}} wormhole. The model is solvable, which allows studying
the properties of the quantum states. In particular   analysing in
detail the matching conditions of quantum states at the throat one
finds that entangled states are generated  through the
non-trivial topology of the wormhole, which in turn allows arguing that
entanglement will be robust under perturbations of the actual shape of
the connecting tube. 
 
We expect that our model can be realised in the laboratory studying
two sheets of bilayer graphene joined by a bilayer nanotube. While
unable to describe low energy localised states, our equations will be
valid in the regime of intermediate energy and for weak enough
curvature. For bilayer graphene  the entanglement properties that we
find have a clear interpretation in terms of geometrical properties of
the microscopical lattice. Curved bilayer graphene has been recently
studied in \cite{Curvatronics2016,FewLayers2018}, where equations have
been written that describe the appropriate mixture of pseudospin
states in the curved case. In particular it has been shown that, if
the energy is higher compared to the angular momentum, but low enough
for the excitations  to still be in the non-relativistic regime, then the
solutions are well approximated by certain types of L\'evy-Leblond
non-relativistic spinors. There is a large literature describing the effects of strain and curvature in monolayer graphene through the Dirac equation, see for example  \cite{Villareal2017,Vozmediano2007_smooth_ripples,Vozmediano2013,Voz2013,Settnes2016,Iorio2015,Lewenkopf2015,Yang2015}, as well as on other effects such as those due to dislocations, or strain in general \cite{Chernodub2017,Peeters2013,Peeters2013_2,CastroNeto2014,Zhu2014}.

The idea of experimentally 
  studying curved sheets of
  monolayer graphene to study  analogues of quantum fields in the vicinity
  of event horizons is not new 
 (see \cite{Cvetic:2012vg} and references therein).

In our case, we focus our
attention on bilayer graphene since strain can be applied to it both
along its surface and perpendicularly, so that in principle it has a
greater potential of generating interesting electronic behaviour with
respect to the monolayer. Our considerations on the role of strain are
mitigated by the fact that the curved bilayer equations that we use
apply to the case of mild curvature. Regardless of this, the main
original point that we make is that by using condensed matter
analogues of wormhole geometries it should be possible to construct entangled states. 
 
\section{The model} 
    
\subsection{Wormhole geometry} 
The Bronnikov-Ellis wormhole geometry
\textcolor{red}{\cite{Bronnikov:1973fh,Ellis:1973yv}}  is 
\ben
ds_{4,BE}^2 = -dt^2 + dr^2 + (r^2 +a^2) d \Omega_2^2 \, , \label{BEW} 
\een
where $d\Omega_2^2$ is the metric on the $2$-sphere. The surface at $\theta =\pi/2$, $t=0$  has two flat ends joined by an Einstein-Rosen bridge: 
\ben 
ds_2^2 = dr^2 + C^2(r)  d\varphi^2 \, ,  \label{eq:metric_2D}  
\een 
where 
\ben 
C(r) = \sqrt{r^2 +a^2} \, , \qquad r \ge 0 \, . 
\een 
This is the metric we will use for our model. The points with $r=0$ constitute a circle of radius $a$ and correspond to the innermost part of wormhole's throat. 
It is useful to visualise it by embedding it in $3$-dimensional
Euclidean space with coordinates $\{x,y,z\}$: setting 
\bea 
C^2(r) &=& x^2 + y^2 = \rho^2\, , \qquad \rho \ge a \, , \\ 
dr^2 &=& dz^2 + d\rho^2 \, ,  
\eea 
we obtain 
\ben 
z = \pm a \cosh^{-1}\left( \frac{\rho}{a} \right) \, .  \label{eq:embedding}
\een 
Thus we can cover the wormhole using two coordinate patches $\{ r_1, \varphi_1\}$, $\{r_2, \varphi_2 \}$, one for each sign of the $z$ variable. 

As pointed out in \cite{Gibbons:2011rh}, the
the surface defined by (\ref{eq:embedding}) is a catenoid, the shape
taken up by a circularly symmetric soap-film.
Because  the spacetime metric (\ref{BEW})
is \emph{ultra-static}, i.e. $g_{tt}=-1$,
semi-classical scattering by the wormhole is given by
geodesics on the catenoid which may integrated  in   terms
of elliptic  functions (\ref{eq:embedding}). This contrasts
with black hole metrics for which $g_{tt} $
depends on the spatial coordinate and changes
sign on the event horizon. For massless excitations
it had been suggested that one should use
an isometric embedding of the so-called optical metric,
i.e. the spatial metric divided by $|g_{tt}|$.
This may be isometrically embedded into $3$-dimensional Euclidean space
but as was pointed out in \cite{Cvetic:2012vg} the
embedding must always exclude  a neighbourhood of the
event horizon. For example in the case of the
Schwarzschild black hole, Flamm's embedding as a paraboloid
of revolution may be extended to cover the entire
Cauchy surface. However the embedding of the optical metric
gets no closer to the event event horizon at $R_S=2M$, where
$R_S$ is the Schwarzschild area coordinate and $M$ the mass,
than a factor of $\frac{9}{4}M$ \cite{Gibbons:2011rh}.   

The two-geometry has negative curvature, which Ricci scalar 
\ben 
R^\mu{}_\mu = - 2 \frac{C^{\prime\prime}}{C} = - \frac{2 a^2}{(r^2+a^2)^2} \, . 
\een 
The maximum curvature is at the throat $r=0$ with value $-\frac{1}{2a^2}$. Therefore in the limit of large $a$ it can be made arbitrarily small.  
 
Every two-dimensional metric is conformally flat, and in fact we can introduce a new coordinate $R$ and write \eqref{eq:metric_2D} as 
\ben 
ds_2^2 = \frac{C(r)^2}{R^2} \left( dR^2 + R^2 d\varphi^2 \right) \, , \qquad \frac{dr^2}{C^2(r)} = \frac{dR^2}{R^2} \, .  \label{eq:metric_2D_R}
\een 
Integrating with the initial condition $\left. R\right|_{r=0} = a$ we find 
\ben 
r = \frac{1}{2} \left( R - \frac{a^2}{R} \right) \, . 
\een 
The form \eqref{eq:metric_2D_R} of the metric is useful as the $R$, $\varphi$ variables are global coordinates for the wormhole. In fact if we introduce cartesian variables $X$, $Y$ such that $dR^2 + R^2 d\varphi^2 = dX^2 + dY^2$, then the inversion transformation in a circle of radius $a$, $\iota : (X,Y) \mapsto (X^\prime, Y^\prime )$ with 
\ben 
\left( \begin{array}{c} X ^\prime \\ Y^\prime \end{array} \right) = \frac{a^2}{X^2+Y^2} \left( \begin{array}{c} X  \\ Y \end{array} \right) \, , \label{eq:inversion} 
\een 
is an isometry of metric such that $\iota (r) = - r$. Therefore the surface $R \ge a$ maps to the upper part of the wormhole $z\ge 0$, and the punctured disc $0< R \le a$ maps to lower part $z \le 0$. The punctured plane $\mathbb{R} \setminus \{ 0 \}$ covers the wormhole. 
 
\begin{figure}
\includegraphics[height=0.25\textwidth,width=0.45\textwidth]{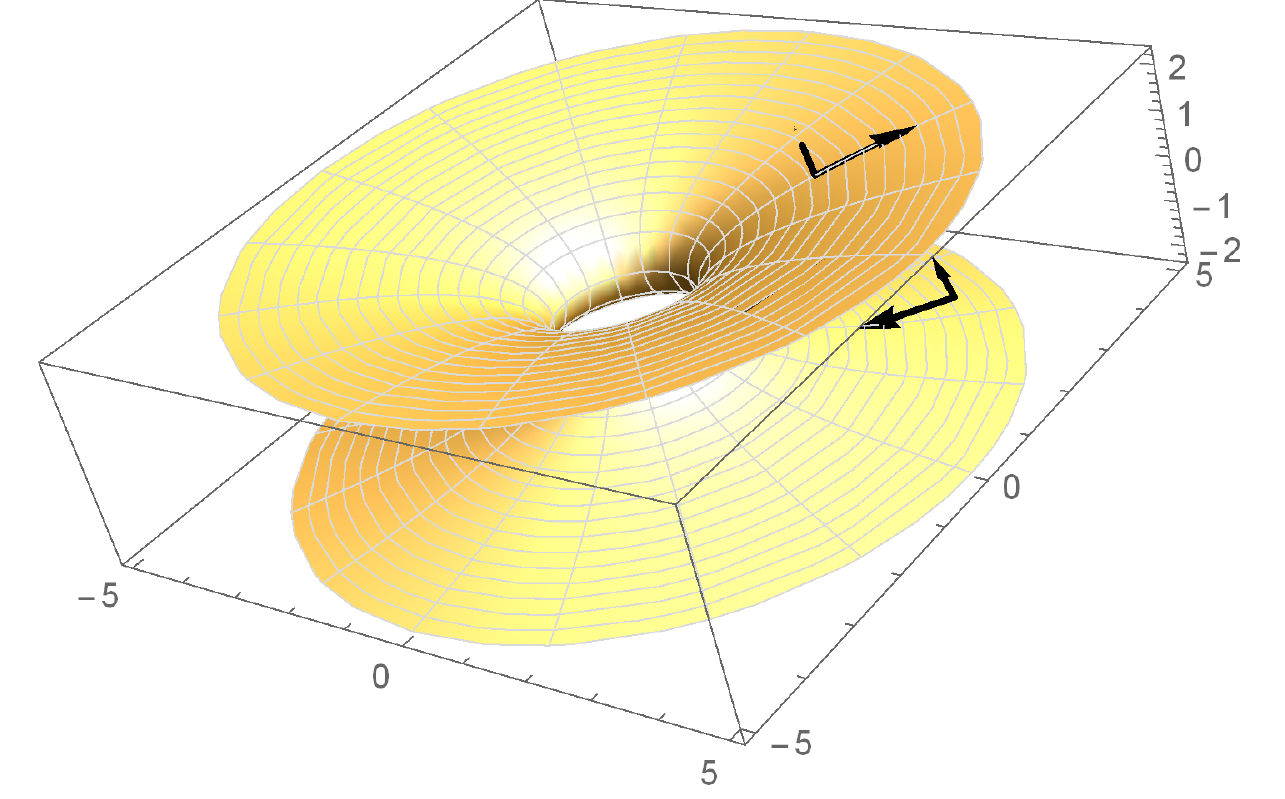} 
 \caption{\label{fig:switch}An orthonormal basis is parallelly propagated through the throat, first by decreasing r until $r=0$, then changing into the increasing $r^\prime = - r$ coordinate. The chirality of the base smoothly changes from that in the upper sheet to that in the lower sheet.}
  \end{figure}

The sign change of $r$ crossing the the wormhole's throat is significant. Consider the vielbeins $e^r = \frac{C}{R} dR$, $e^\varphi = C d\varphi$. Under the inversion \eqref{eq:inversion} they change as: $\iota (e^r) = - e^r$, $\iota (e^\varphi) = e^\varphi$. This effect can be understood following the  transport around a $\varphi = \text{const.}$ line of the vielbein basis across the wormhole's throat, as depicted in fig.\ref{fig:switch}. We interpret this by saying that the wormhole acts as a \textit{chirality switch}, changing smoothly the chirality of a basis from the upper sheet to the lower sheet. 

%
%
 
\subsection{L\'evy-Leblond fermions and the Eisenhart-Duval lift} 
Non-relativistic fermions have been studied originally by
L\'evy-Leblond \textcolor{red}{\cite{LevyLeblond:1967zz}},
and in a curved geometry they are described
by two time-dependent spinors $\chi_1$, $\chi_2$ with two components
which satisfy the following set of coupled, first order equations
\bea \label{eq:LevyLeblond1} &&  i \hbar v_0 D \chi_1 + \mathcal{E}
\chi_2  = 0 \, , \\  && 2 m v_0 \chi_1 + i \hbar D \chi_2  = 0 \,
, \label{eq:LevyLeblond2} \eea  where $D =  \sigma^\mu \,
\nabla_\mu^{(2)}$ is the $2$--dimensional Dirac operator, the
$\sigma^\mu$ are the curved Pauli matrices, $\mathcal{E}$ is the
energy, $m$ is the mass of the fermion and $v_0$ a characteristic
speed. In the original work of L\'evy-Leblond $v_0=c$ the speed of
light, and the equations above were obtain as low energy limit of the Dirac
equation, displaying Galilean symmetry in the flat case in the absence
of interaction; however in the case of bilayer graphene $v_0 = v_F \sim 10^6 ms^{-1}$, the Fermi velocity in
graphene.  The equations admits two independent solutions
corresponding to propagating degrees of freedom for the spin. 
 
It is a known fact \cite{Horvathy:1993fd,Horvathy1996} that the L\'evy-Leblond equations in the metric \eqref{eq:metric_2D_R} are equivalent to the massless Dirac equation in the $1+3$-dimensional Eisenhart-lift metric 
\bea \label{eq:metric_Eisenhart}
ds_{4,Eis}^2 &=& -v_F^2 \, dT^2 + dZ^2 + \frac{C(r)^2}{R^2} \left( dR^2 + R^2 d\varphi^2 \right) \nn \\ 
&=& 2 dU dV +  \frac{C(r)^2}{R^2} \left( dR^2 + R^2 d\varphi^2 \right) \, , \label{eq:4D_metric}
\eea 
where $T$ and $Z$ are new coordinates. The coordinates $U$, $V$ are defined as 
\ben 
U = \frac{Z + v_0 T}{\sqrt{2}} \, , \qquad V = \frac{Z - v_0 T}{\sqrt{2}} \, . 
\een 
The equivalence works as follows. We begin with a $4$-dimensional spinor $\psi$ with $4$ components that satisfies 
\ben \label{eq:massless_Dirac}
\Gamma^\mu \nabla_\mu \psi = 0 \, , 
\een 
where $\Gamma^\mu$ and $\nabla_\mu$ are Gamma matrices and spinorial covariant derivatives in $4$ dimensions. The $4$D metric admits a null Killing vector $\pa_V$, so we can look for specific solutions of the form 
\ben \label{eq:4D_spinor_v_factor}
\Psi (U, V, r, \varphi) = e^{i \frac{m v_0}{\hbar} V} e^{-i \frac{\mathcal{E} U}{\hbar v_0}} \left( \begin{array}{c} \chi_1 ( r, \varphi) \\ \chi_2 (r, \varphi) \end{array} \right) \, . 
\een 
Then one can define Gamma matrices adapted to the null form of the metric, and re-write \eqref{eq:massless_Dirac} in the form (\ref{eq:LevyLeblond1}-\ref{eq:LevyLeblond2}) with the identification $U =  v_0 \, t$ \cite{Cariglia2012,Curvatronics2016}. 
 
We will solve the four-dimensional null Dirac equation using the following ansatz: 
\ben 
\Psi = \psi_1(U,V) \otimes \psi_2 (R,\varphi) \, , \label{eq:psi1}
\een 
which is adapted to the form of the metric \eqref{eq:4D_metric}. Here the $\psi_i$, $i=1,2$, are two-dimensional spinors with two components each, and the Dirac operators decomposes as a sum $D + \mathcal{D}$, where $D$ has been defined above as the Dirac operator on the wormhole metric, and $\mathcal{D}$ is the Dirac operator on flat two-dimensional Minkowski space with metric $2 dU dV$. We use the following representation for the flat Gamma matrices associated to the variables in the order $\{U,V,R,\varphi\}$: 
\ben 
\Gamma^1 = \left( \begin{array}{cc} 0 & 1 \\ 0 & 0 \end{array} \right) \otimes \mathbb{I}  \, , \qquad \Gamma^2 = \left( \begin{array}{cc} 0 & 0 \\ 2 & 0 \end{array} \right) \otimes \mathbb{I} \, , \qquad \Gamma^3 = \sigma_3 \otimes \sigma_1 \, , \qquad \Gamma^4 = \sigma_3 \otimes \sigma_2 \, . 
\een 
We look for solutions such that 
\ben \label{eq:split_solution_1}
\mathcal{D} \psi_1 = i k \sigma_3 \psi_1 \, , \qquad k \in \mathbb{R} \, ,  
\een 
finding 
\ben 
\psi_1 = e^{i \frac{mv_0}{\hbar} V} e^{- i \frac{\mathcal{E}}{\hbar v_0}U} 
\left( \begin{array}{c} s \sqrt{\frac{\mathcal{E}}{2m v_0^2}} \\ - 1 \end{array} \right) \, , 
\een 
where $s = \text{sgn}(k) = \text{sgn}(\mathcal{E}) = \pm 1$, $\mathcal{E}$ has the interpretation of energy of the L\'evy-Leblond fermion and the eigenvalue $k$ is given by $k = s \frac{\sqrt{2m\mathcal{E}}}{\hbar}$. 
 
Next, inserting \eqref{eq:split_solution_1} into the four-dimensional massless Dirac equation we obtain 
\ben 
D \, \psi_2 = - i k \, \psi_2 \, . \label{eq:D_psi2_first_form}
\een 
This equation can be solved using the properties of the Dirac equation under conformal transformations. Let $\tilde{D}$ be the flat Dirac operator associated to the metric $dR^2 + R^2 d\varphi^2$, and set $\Omega = \frac{C}{R}$. Then it is a well known fact \cite{Gary1982,Gibbons:1981ux,Gibbons:2008rs} that 
\ben 
D\, \psi_2 = \Omega^{-\frac{3}{2}} \tilde{D} \, \tilde{\psi}_2 \, , \qquad \psi_2 = \Omega^{-\frac{1}{2}} \, \tilde{\psi}_2 \, ,  
\een 
and  eq.\eqref{eq:D_psi2_first_form} becomes $\tilde{D} \, \tilde{\psi}_2 = - i k \frac{C}{R} \tilde{\psi}_2$. Parameterizing $\tilde{\psi}_2 = e^{i \frac{p_\varphi \varphi}{\hbar}} \left( \begin{array}{c} \tilde{\alpha}(R) \\ \tilde{\beta}(R) \end{array} \right)$, with $p_\varphi = \hbar \frac{2l+1}{2}$, $l \in \mathbb{Z}$, we obtain the coupled equations 
\bea 
\left( \pa_R + \frac{1}{2R} + \frac{p_\varphi}{\hbar R} \right) \tilde{\beta} &=& - i k \frac{C}{R} \tilde{\alpha} \, , \label{eq:alpha_tilde} \\ 
\left( \pa_R + \frac{1}{2R} - i \frac{p_\varphi}{\hbar R} \right) \tilde{\alpha} &=& - i k \frac{C}{R} \tilde{\beta} \, . \label{eq:beta_tilde}
\eea 
We will discuss their solutions in more detail in the next session. The complete form of the solution of the four-dimensional massless Dirac equation is:  
\ben \label{eq:4d_solution_tensor_product}
\Psi_{m,\mathcal{E},k} = \sqrt{\frac{R}{C}} e^{i \frac{mv_0}{\hbar} V} e^{i \frac{\mathcal{E}}{\hbar v_0}U} e^{i p_\varphi \varphi} 
\left( \begin{array}{c} s \sqrt{\frac{\mathcal{E}}{2m v_0^2}} \\ - 1 \end{array} \right) \otimes \left( \begin{array}{c} \tilde{\alpha}(R;k) \\ \tilde{\beta}(R;k) \end{array} \right) \, . 
\een 

We conclude this section discussing the properties of our solutions under parity. The chirality matrix 
\ben 
\Gamma^* =  \sigma_3 \otimes \sigma_3 = 
\left(  \begin{array}{cccc} 
	1 & 0 & 0 & 0 \\ 0 & -1 & 0 & 0 \\ 0 & 0 & -1 & 0 \\ 0 & 0 & 0 & 1 
	  \end{array} \right) \, , 
\een 
commutes with the four-dimensional Dirac operator. Therefore our solutions must decompose into eigenspinors with even and odd values of chirality, $\Gamma^* \Psi^\pm = \pm \Psi^\pm$. In \cite{FewLayers2018} we have shown how the $4$ dimensional chirality is directly related to the isospin of the L\'evy-Leblond fermions in 2D, and that the two isospin states correspond to electronic states that are localised in either of the two graphene planes. Physical states in bilayer graphene, however, present a well defined mixture of pseudospin. 
 
We achieve such  decomposition inthe following way. If $\tilde{\alpha}(R;k)$, $\tilde{\beta}(R;k)$ are solutions of (\ref{eq:alpha_tilde}-\ref{eq:beta_tilde}), then $\tilde{\alpha}(R;-k)= - \tilde{\alpha}(R;k)$, $\tilde{\beta}(R;-k) = \tilde{\beta}(R;k)$ solve the equations for $k \rightarrow - k$. Then $\tilde{\alpha}(R;k) = s \tilde{\alpha}(R;|k|)$, $\tilde{\beta}(R;k) = \tilde{\beta}(R;|k|)$. Therefore taking linear combinations of \eqref{eq:4d_solution_tensor_product} we can construct solutions of opposite chirality: 
\bea 
\Psi_{m,\mathcal{E},k}^+ &=& \sqrt{\frac{R}{C}} e^{i \frac{mv_0}{\hbar} V} e^{i \frac{\mathcal{E}}{\hbar v_0}U} e^{i p_\varphi \varphi} 
\left( \begin{array}{c} s \sqrt{\frac{\mathcal{E}}{2m v_0^2}} \tilde{\alpha}(R;|k|) \\ 0 \\ 0 \\   - \tilde{\beta}(R;|k|) \end{array} \right) \, , \\ 
\Psi_{m,\mathcal{E},k}^- &=& \sqrt{\frac{R}{C}} e^{i \frac{mv_0}{\hbar} V} e^{i \frac{\mathcal{E}}{\hbar v_0}U} e^{i p_\varphi \varphi} 
\left( \begin{array}{c} 0 \\ s \sqrt{\frac{\mathcal{E}}{2m v_0^2}} \tilde{\beta}(R;|k|) \\   - \tilde{\alpha}(R;|k|) \\ 0 \end{array} \right) \, .  
\eea

\section{The matching conditions and entanglement\label{sec:matching}} 
In this section we analyze the matching conditions at $r=0$. In the previous section we have found the global form of the spinor $\Psi$ over the whole wormhole, however in the $R$ picture the two sides of the wormhole are related non-linearly by the inversion $\iota$. In this section we will express our solution in the $r$ picture, such that the two sides of the wormhole will be more easily comparable. 
 
Consider the radial equations (\ref{eq:alpha_tilde}-\ref{eq:beta_tilde}). Remembering that $\tilde{\alpha} = \sqrt{\frac{C}{R}} \alpha$, and similarly for $\beta$, and expressing everything in terms of the $r$ variable, the radial equations become 
\bea 
\mathcal{O}^\dagger \alpha(r;|k|) &=& i |k| \beta(r;|k|)  \, , \label{eq:curved_coupled1_susy}  \\ 
\mathcal{O} \beta(r;|k|) &=&  - i |k| \alpha(r;|k|) \, ,   \label{eq:curved_coupled2_susy}  
\eea
where the operators $\mathcal{O}$ and $\mathcal{O}^\dagger$, mutually adjoint relative to the measure $C(r) dr$, are given by 
\bea 
\mathcal{O} &=& \left(  \pa_r + \frac{C^\prime}{2C} +  \frac{p_\varphi}{\hbar C} \right)  \, , \label{eq:O} \\ 
\mathcal{O}^\dagger &=& \left( -  \pa_r - \frac{C^\prime}{2C} +  \frac{p_\varphi}{\hbar C} \right)  \, ,  \label{eq:Odag}
\eea 
and $C'(r) = \frac{dC}{dr}$. The equations can be separated by deriving once more, with the result 
\bea 
\mathcal{O}  \mathcal{O}^\dagger \, \alpha &=& 2 \frac{m \mathcal{E}}{\hbar^2} \alpha \, ,  \label{eq:OdagO_lambda1} \\ 
\mathcal{O}^\dagger \mathcal{O} \, \beta &=& 2 \frac{m \mathcal{E}}{\hbar^2} \beta \,.  
\eea 
Since both $\mathcal{O}  \mathcal{O}^\dagger$ and $\mathcal{O}^\dagger \mathcal{O}$ are positive operators, then $\mathcal{E} \ge 0$. The equations are supersymmetric since defining  
\ben 
\mathcal{Q} = \left( \begin{array}{cc} 0 & \mathcal{O} \\ 0 & 0 \end{array} \right) \, , \qquad 
\mathcal{Q}^\dagger = \left( \begin{array}{cc} 0 & 0 \\ \mathcal{O}^\dagger  & 0 \end{array} \right) \, , 
\een
yields  
\ben 
\left\{ \mathcal{Q} , \mathcal{Q}^\dagger  \right\} = 2 \frac{m \mathcal{E}}{\hbar^2} \mathbb{I} \, . 
\een

The ground state wavefunction, $\mathcal{E}=0$, satisfies 
\ben 
\alpha = \frac{1}{(r^2+a^2)^{\frac{1}{4}}} \left(r+\sqrt{r^2+a^2}\right)^{\frac{2l+1}{2}} \, , \qquad \beta = \frac{1}{(r^2+a^2)^{\frac{1}{4}}} \left(r+\sqrt{r^2+a^2}\right)^{-\frac{2l+1}{2}} \, , 
\een 
and is not normalisable for any value of $l$. 

Recall that the radial vierbein $e^r$ changes sign when crossing the wormhole's throat, and that the radial variable $r$ also changes sign In dealing with spinors, we need to include a unitary change of basis for the Gamma matrices. This is realised by the unitary transformation 
\ben \label{eq:Gamma_change_of_basis}
\Gamma^\mu_{down}(z = 0^-)  = \hat{U} \, \Gamma^\mu_{up}(z=0^+) \, \hat{U}^\dagger  \, , 
\een 
where "up" and "down refer to portions of the hyperboloid with $z>0$ and, respectively, $z<0$. The matrix is given by 
\ben \label{eq:transition_matrix} 
\hat{U} =  \mathbb{I} \otimes \sigma_2 =  
\left( \begin{array}{cccc} 
	0 &- i & 0 & 0 \\ 
	i & 0 & 0 & 0 \\ 
	0 & 0 & 0 & -i \\ 
	0 & 0 & i & 0 
\end{array} \right) \, ,  
\een 
and is such that the operation \eqref{eq:Gamma_change_of_basis} changes the sign of $\sigma_3 \otimes \sigma_1$, which is the Gamma matrix associated to $e^r$, and leaves the other Gamma matrices unchanged. 
Spinors transform at the junction as 
\ben 
\psi_{down}(z=0^-) = \hat{U} \psi_{up} (z=0^+) \, ,  
\een 
and after the transition they must satisfy, relative to the new variable $r^\prime$, the same equations of motion. Summarizing, in the $r$-picture the two independent solutions of the null Dirac equation in the full geometry should be given by: 
\ben 
\Psi^{\pm} = \left\{ \begin{array}{cl} \Psi^{\pm} (r) \, ,  \quad  r>0 \, , \\ \hat{U} \Psi^{\pm} (r^\prime) \, , \quad  r^\prime > 0 \, , \end{array} \right. 
\een 
with $r^\prime = - r$. An explicit check shows that these are correct solutions. For example the explicit form of $\hat{U} \Psi^+ (r^\prime)$ is 
\ben 
U \Psi^+ (r^\prime) =  \, e^{i \frac{mv_0}{\hbar} V} e^{-i \frac{\mathcal{E}}{\hbar v_0}U} e^{i p_\varphi \varphi} 
\left( \begin{array}{c} 0 \\ i s \sqrt{\frac{\mathcal{E}}{2m v_0^2}} {\alpha}(r^\prime;|k|) \\  i  {\beta}(r^\prime;|k|) \\ 0 \end{array}  \right) \, .   
\een 
This is the same as a $\Psi^-(r^\prime)$ spinor if we make the identification $\beta^\prime (r^\prime) = i \alpha (r^\prime)$, $\alpha^\prime (r^\prime) = - i \beta (r^\prime)$, the identification being made possible by the fact that  changing $r$ into $r^\prime$ interchanges the operators $\mathcal{O}$ and $\mathcal{O}^\dagger$: 
\ben 
\mathcal{O}_{r^\prime} = \mathcal{O}^\dagger_r \, , \qquad \mathcal{O}^\dagger_{r^\prime} = \mathcal{O}_r \, .  
\een 
A similar consideration holds for  $\hat{U} \Psi_- (r^\prime)$, which looks like a positive chirality solution in the lower hyperboloid. 
 
Far away from the throat $C(r) \sim r$ and our solutions look like solutions of the Dirac equation in flat space, however whenever on the upper sheet there is a positive chirality solution, in the lower sheet there is a lower chirality solution. In other words our full solutions behave asymptotically as entangled states: if we describe asymptotic states using states from the product Hilbert space of two flat sheets, then we can write
\bea 
lim_{|r| \rightarrow + \infty} \Psi^+ &=& \psi^+(r) \otimes \psi^-(r^\prime) \, , \\ 
lim_{|r| \rightarrow + \infty} \Psi^- &=& \psi^-(r) \otimes \psi^+(r^\prime)  \, . 
\eea 
At large $r$ one sees perfectly anticorrelated entangled pairs. 
 
The main reason why the results work is that there is a glueing condition at $z=0$, rather than the specific form of the throat. Therefore even for metrics that are not axisymmetric and for more complicated forms of the wormhole the same type of entangled solutions should appear: we expect the result to be robust under deformations of the surface, as it is of a topological nature.

\section{Condensed matter analog with bilayer graphene}   
The $2$ dimensional surface and the L\'evy-Leblond fermions we have discussed are expected to have a condensed matter analogue that can be built using bilayer graphene. As discussed originally in \cite{Curvatronics2016}, and then refined in \cite{FewLayers2018}, certain types of L\'evy-Leblond fermions are expected to arise as the electronic excitations of bilayer graphene for a surface with mild curvature. Physically the surface can be produced by joining two separate graphene bilayers by a bilayer nanotube. In the literature the similar case with monolayer graphene has been discussed from various points of view \cite{GonzalezHerrero2010,Atanasov2011,PincakSmotlacha2013,Capozziello2017}, although not from the point of view of entanglement, neither from the point of view of a bilayer with its increased sensitivity to strain. It is well known that in order to join a nanotube to single graphene sheet it is necessary to introduce heptagonal defects \cite{GonzalezHerrero2010}, and that these have the distinct signature of producing localised states at the bottom of the parabolic energy bands. The effect of heptagons can be modelled using non-abelian gauge fields that mix excitations around the two different Dirac points of single layer graphene. Such feature is not included in our model: in other words, we consider energies that are large enough to make the contribution from the gauge fields negligible. 
 
We expect our model to work under a set of approximations that correspond to a well defined subset in the possible parameter space of the theory. Four main approximations are needed: low energy, low curvature, high energy to angular momentum, and high energy relative to the flux of the effective gauge fields. We describe these approximations one by one. 
 
First of all, the energy of the excitation should be smaller than $\gamma\sim 0.4eV$ , the hopping parameter between different layers, otherwise the dispersion relation is no longer quadratic in momenta and the notion of non-relativistic fermions breaks. Second, the curvature makes a contribution to the local energy of the states. Let the average radius of curvature $\bar{\rho}$ be defined from the curvature scalar as $\left| R^\mu{}_\mu \right|=  \frac{2}{\bar{\rho}^2}$. Then, as discussed in \cite{Curvatronics2016}, it is necessary that  $|\bar{\rho}| >> 1,16nm$ in order that the correction to the energy due to the curvature is not so big as to make invalid the approximation that the electronic states are close to the Dirac points. In our case the minimum value of the average radius of curvature can be obtain from $|R^\mu{}_\mu|_{max} = \frac{1}{2a^2} = \frac{2}{\bar{\rho}_{min}^2}$, so that we need 
\ben \label{eq:a_min}
2 a >> 1.16nm \, . 
\een 
 In addition to this, the theory developed in \cite{Curvatronics2016,FewLayers2018} assumes that the curvature is sufficiently smooth so that in first approximation the distance between the two graphene layers is unchanged, and the stacking remains of type AB. This will not be rigorously true for curved surfaces, but it will be a good approximation if the radius of curvature is much bigger than the lattice scale. The bound obtained above corresponds to a minimum radius of curvature of the order of $4$ times the carbon-carbon distance, so it is of the right order of magnitude. 
 
Third, in \cite{FewLayers2018} it has been shown that the electronic states do not in general correspond to L\'evy-Leblond fermions. Only one specific combination of isospin is allowed, and this never rigorously corresponds to L\'evy-Leblond excitations. However, if the angular momentum is small compared to the energy then the physical states are well approximated by L\'evy-Leblond solutions: in the present case we find the following bound on the energy 
\ben 
\frac{4}{3\sqrt{3} a^2} \frac{\hbar^2 v_F^2}{\gamma} |\frac{2l+1}{2}| << |\mathcal{E}| \, .  
\een 
If we substitute the physical values in the equation above and choose the minimum admissible value for $a$ according to \eqref{eq:a_min} then we get $|\mathcal{E}| >> (2.48 eV) |l+ \frac{1}{2}|$, which is  bigger than $\gamma$. Instead we consider a higher window of values of $a$, asking that for a number $p\in \mathbb{R}$ it should be $a >> p\, nm$. Then we find  
\ben 
\left(\frac{0.83}{p^2}eV\right) |\frac{2l+1}{2}| << |\mathcal{E}| \, . \label{eq:energy_lower_bound}
\een 
 
The last approximation we need is that the flux term due to the effective gauge fields at the heptagons is negligible, and this leads to the requirement $|\frac{2l+1}{2}| >> \frac{3}{2}$.  in eq.\eqref{eq:energy_lower_bound}. So we need to consider values $a>>2nm$ to obtain a window of valid energies below $\gamma$. To summarise, we are therefore in the following window of intermediate energies  
\ben \label{eq:energy_window}
\left(\frac{0.83}{p^2}eV\right) |\frac{2l+1}{2}| << \mathcal{E} << \gamma \, , \qquad |\frac{2l+1}{2}| >> \frac{3}{2}
\een

The L\'evy-Leblond approximated solutions fund in \cite{FewLayers2018} were written in terms of the spinors $\chi_1$, $\chi_2$ of eqs.(\ref{eq:LevyLeblond1}-\ref{eq:LevyLeblond2}), and are such that for positive energy $\sigma_1 \chi_1 = - \chi_1$, and for negative energy $\sigma_1 \chi_1 = + \chi_1$. In this limit of high energy to angular momentum we obtain the L\'evy-Leblond approximated solutions  in the top part of the hyperboloid,  
\ben \label{eq:LL_bilayer_graphene}
\psi_{\mathcal{E}} =  \, e^{i \frac{mv_0}{\hbar} V} e^{-i \frac{\mathcal{E}}{\hbar v_0}U} e^{i p_\varphi \varphi} \left( 
	\begin{array}{c} 
	s  {\alpha}(r;|k|)  \\ -   {\alpha}(r;|k|) \\ s \frac{i\hbar v_F}{\mathcal{E}} \mathcal{O} {\alpha}(r;|k|) \\   \frac{i\hbar v_F}{\mathcal{E}} \mathcal{O}^\dagger {\alpha}(r;|k|) 
	\end{array} \right) \, , 
\qquad z>0 \, . 
\een 
By taking a sum or a difference of the states above with opposite values of $\mathcal{E}$ it is possible to obtain states which have either the first and third component equal to zero, or the second and fourth. Such states were discussed in \cite{FewLayers2018} where they where called `checkered states': they correspond to microscopic configurations where electrons are localized only on atoms of $A$ type on both planes of graphene, or only on atoms of $B$ type. According to the results of sec.\ref{sec:matching} going through the wormhole changes one type of checkered state into the other via multiplication times the $\hat{U}$ matrix. At the microscopical level this is understood in terms of the chirality switch: going through the wormhole changes the orientation of the sheet and this corresponds to changing atoms of type A with atoms of type B. For example a positive energy state $\psi_{\mathcal{E}}$ can be decomposed in terms of checkered eigenstates $\psi_A$, $\psi_B$: 
\ben 
\psi_{\mathcal{E}} = \frac{1}{\sqrt{2}} \psi_A - \frac{1}{\sqrt{2}} \psi_B \, , 
\een 
where the meaning of $A$, $B$ can be defined asymtpotically away from the throat. 
If one is able to project the energy eigenstate into a checkered eigenstate, then the resulting state will asymptotically look like a pair of entangled particles on the two sheets, with opposite checkering. This can be verified experimentally by STM scannening tunneling spettroscopy which can provide atomic resolution \cite{STM2012}.

\begin{figure}
\includegraphics[height=0.25\textwidth,width=0.9\textwidth]{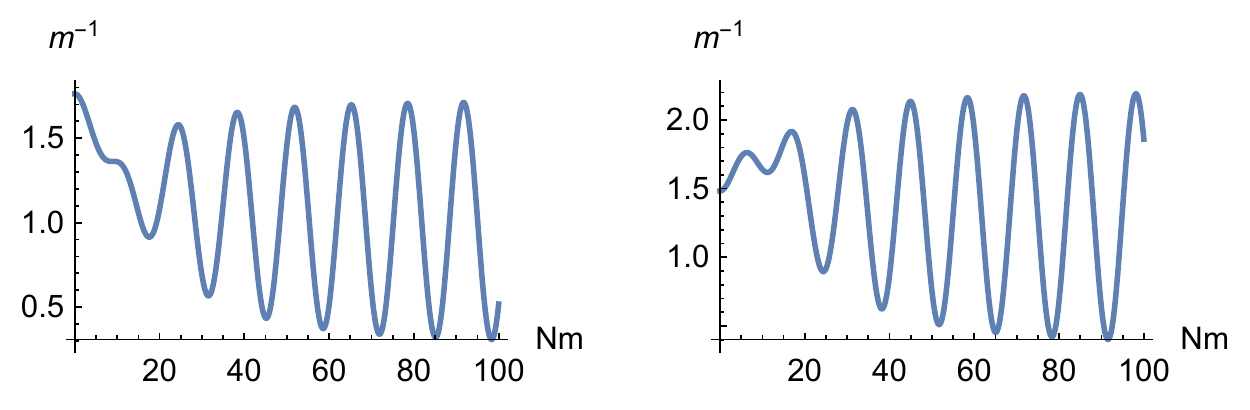} 
 \caption{\label{fig:l_3}The function $\rho(r)$ for $l=3$ and for two different boundary conditions for the function $f$. On the left we chose $f(0) \neq 0$, $f^\prime (0)=0$, and on the right $f(0) = 0$, $f^\prime (0)\neq 0$ A value of the energy of $0.06eV$ has been used. The vertical amplitude is illustrative as the wavefunctions are not normalized.}
  \end{figure}

\begin{figure}
\includegraphics[height=0.25\textwidth,width=0.9\textwidth]{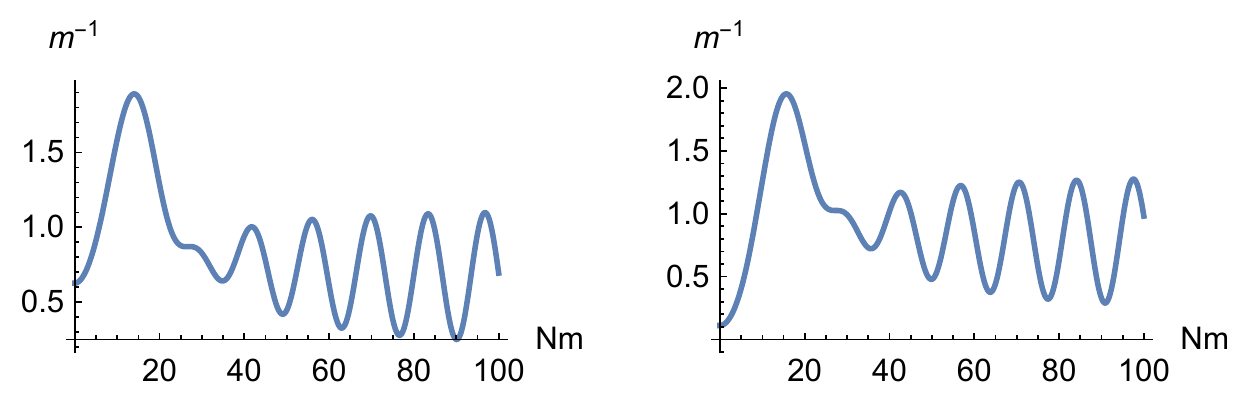} 
 \caption{\label{fig:l_5}The function $\rho(r)$ for $l=5$ and for two different boundary conditions for the function $f$. On the left we chose $f(0) \neq 0$, $f^\prime (0)=0$, and on the right $f(0) = 0$, $f^\prime (0)\neq 0$ A value of the energy of $0.06eV$ has been used. The vertical amplitude is illustrative as the wavefunctions are not normalized.}
  \end{figure}

We now analyse numerically some of the explicit solutions of \eqref{eq:LL_bilayer_graphene}, where $\alpha$ satisfies \eqref{eq:OdagO_lambda1}. Fixed parameters in the system are the maximum energy $\gamma \sim 0.4eV$, and the effective mass $m = \frac{\gamma}{2v_F^2} \sim 3.2 \cdot 10^{-32}kg$. We choose a relatively large value for the throat parameter $a = 20nm$, so that the lower bound in energies in \eqref{eq:energy_window} becomes $0.007eV$ for $l=3$ and $0.011eV$ for $l=5$. We analyse the spatial density of states which is given by 
\ben 
\rho(r) = \sqrt{-g} \, \bar{\psi}_{\mathcal{E}} \Gamma^T \psi_{\mathcal{E}} = \frac{C(r)}{v_F} \psi^\dagger_{\mathcal{E}} (r) \psi_{\mathcal{E}} (r) \, .  
\een 
Figures \ref{fig:l_3} and \ref{fig:l_5} display the spatial density as a function of $r$ for the cases $l=3$ and, respectively, $l=5$. In both cases the density becomes periodic asymptotically, however the presence of the throat is clearly visible in the range $r \le a$. 
 
\section{Conclusions} 
In this work we have proposed a simple, integrable model of entanglement between fermions that is generated from geometrical and topological properties of the underlying spacetime. From the point of view of  General Relativity the objective was to have an example with a clear mechanism and interpretation, rather than necessarily describing a realistic spacetime. We chose the wormhole geometry since it can be implemented in the laboratory via a condensed matter analogue, using two bilayer graphene sheets joined by a bilayer nanotube. We expect that this geometry, if it can be obtained experimentally, can be used to generate states that display the features of entanglement asymptotically on the two sheets. 
 
Our geometrical model is valid for a smooth enough surface and in a well defined window of intermediate energies. A more realistic treatment that takes into account the microscopic lattice, the presence of heptagonal defects, and the physics far from the Dirac points requires a first principles analysis that will be undertaken in future work.

 
\begin{acknowledgments} 
MC was funded by the CNPq under project 303923/2015-6, and by a Pesquisador
Mineiro project n. PPM-00630-17. 
 \end{acknowledgments} 



\begin{thebibliography}{99}

  
\bibitem{Einstein:1935tc}
  A.~Einstein and N.~Rosen, ``The Particle Problem in the General Theory of Relativity'',
  Phys.\ Rev.\  {\bf 48} (1935) 73.  
  DOI:10.1103/PhysRev.48.73. 

\bibitem{Einstein:1935rr}
  A.~Einstein, B.~Podolsky and N.~Rosen,
 ``Can quantum mechanical description of physical reality be considered complete?'', 
  Phys.\ Rev.\  {\bf 47} (1935) 777.
  DOI:10.1103/PhysRev.47.777. 

  
\bibitem{Maldacena:2013xja}
  J.~Maldacena and L.~Susskind, ``Cool horizons for entangled black holes'',
  Fortsch.\ Phys.\  {\bf 61} (2013) 781. DOI:10.1002/prop.201300020, arXiv:1306.0533. 

\bibitem{Maldacena1}
J.~Maldacena ``Entanglement and the Geometry of Spacetime'', The Institute for Advanced Study Newsletter Fall (2013). 


\bibitem{Maldacena:2017axo}
  J.~Maldacena, D.~Stanford and Z.~Yang,
``Diving into traversable wormholes'', 
  Fortsch.\ Phys.\  {\bf 65} (2017)  1700034. 
  DOI:10.1002/prop.201700034, arXiv:1704.05333. 
  
 \bibitem{Maldacena:2018lmt}
  J.~Maldacena and X.~L.~Qi,
 ``Eternal traversable wormhole'', arXiv:1804.00491. 
   
\bibitem{Susskind2016}    
L. Susskind, ``Copenhagen vs everett, teleportation, and ER = EPR'', Fortsch.\ Phys.\ {\bf 64} (2016) 551-64. DOI:10.1002/prop.201600036, arXiv:1604.02589. 


\bibitem{Flamm} L.~ Flamm, ``Beitr\"age zur Einsteinschen Gravitationstheorie''
  Physikalsicshe Seitschrift {\bf 17} (1916) 448-454. English translation: 
  L.~ Flamm `` Republication of:  Contributions to Einstein's Theor'', Gen. Relativ. Gravit   {\bf 47} (2015) 72. DOI 10.1007/s10714-015-1908-2. 

\bibitem{Gibbons} G.~ W.~ Gibbons ``Editorial Note to:   
  Ludwig Flamm, Contributions to Einstein's Theory'', Gen. Relat. Grav. 
    {\bf 47} (2015) 71. DOI 10.1007/s10714-015-1908-3. 

\bibitem{Israel:1976ur}
  W.~Israel, ``Thermo-field dynamics of black holes'',   Phys.\ Lett.\ A {\bf 57} (1976) 107.
  DOI:10.1016/0375-9601(76)90178-X. 


\bibitem{Lee:1985rp}
  T.~D.~Lee, ``Are Black Holes Black Bodies?'',   Nucl.\ Phys.\ B {\bf 264} (1986) 437.
  DOI:10.1016/0550-3213(86)90493-1. 


\bibitem{Bronnikov:1973fh}
  K.~A.~Bronnikov, ``Scalar-tensor theory and scalar charge'',  Acta Phys.\ Polon.\ B {\bf 4} (1973) 251. 

\bibitem{Ellis:1973yv}
  H.~G.~Ellis, ``Ether flow through a drainhole - a particle model in general relativity'',   J.\ Math.\ Phys.\  {\bf 14} (1973) 104.   DOI:10.1063/1.1666161 
   
\bibitem{Curvatronics2016} 
M. Cariglia, R. Giamb\'o, A. Perali, ``Curvature-tuned electronic properties of bilayer graphene in an effective four-dimensional spacetime'', Phys. Rev. B {\bf 95}, 245426 (2017). DOI:10.1103/PhysRevB.95.245426, arXiv:1611.06254. 
 
\bibitem{FewLayers2018} 
M. Cariglia, R. Giamb\'o, A. Perali, \textit{Electronic properties of curved few-layers graphene: a geometrical approach}, Cond. Matt. {\bf 3(2)}, 11 (2018). DOI: 10.3390/condmat3020011, arXiv:1801.04194. 
 
\bibitem{Villareal2017} 
P. Castro-Villarreal, R. Ruiz-Sánchez, \textit{Pseudomagnetic field in curved graphene}, Phys. Rev. B {\bf 95} 125432 (2017). DOI:10.1103/PhysRevB.95.125432, arXiv:1612.04305. 
 
\bibitem{Vozmediano2007_smooth_ripples} 
F. de Juan, A. Cortijo, M. A. Vozmediano, \textit{Charge inhomogeneities due to smooth ripples in graphene sheets}, Phys. Rev. B {\bf 76} 165409 (2007). DOI:10.1103/PhysRevB.76.165409, arXiv:0706.0176. 
 
\bibitem{Vozmediano2013} 
F. de Juan, J. L. Ma\~nes, M. A. Vozmediano, \textit{Gauge fields from strain in graphene}, Phys. Rev. B {\bf 87} 165131 (2013). DOI:10.1103/PhysRevB.87.165131, arXiv:1212.0924. 
 
\bibitem{Voz2013} 
J. L. Manes, F. de Juan, M. Sturla, M. A. Vozmediano, \textit{Generalized effective Hamiltonian for graphene under nonuniform strain}, Phys. Rev. B {\bf 88} 155405 (2013). DOI:10.1103/PhysRevB.88.155405, arXiv:1308.1595. 
  
\bibitem{Settnes2016} A. P. Jauho, \textit{Pseudomagnetic fields and triaxial strain in graphene}, Phys. Rev. B {\bf 93} 035456 (2016). DOI:10.1103/PhysRevB.93.035456, arXiv:1510.07895. 
 
\bibitem{Iorio2015} 
A. Iorio, P. Pais, \textit{Revisiting the gauge fields of strained graphene}, Phys. Rev. D {\bf 92} 125005 (2015). DOI:10.1103/PhysRevD.92.125005, arXiv:1508.00926. 
 
\bibitem{Lewenkopf2015} 
E. Arias, A. R. Hern\'andez, C. Lewenkopf, \textit{Gauge fields in graphene with nonuniform elastic deformations: A quantum field theory approach}, Phys. Rev. B {\bf 92} 245110 (2015). DOI:10.1103/PhysRevB.92.245110 , arXiv:1511.08639. 
 
\bibitem{Yang2015} 
B. Yang, \textit{Dirac cone metric and the origin of the spin connections in monolayer graphene}, Phys. Rev. B {\bf 91} 241403 (2015). DOI:10.1103/PhysRevB.91.241403, arXiv:1402.0941. 
  
\bibitem{Chernodub2017} 
M. N. Chernodub, M. A. Zubkov, \textit{Chiral anomaly in Dirac semimetals due to dislocations}, Phys. Rev. B {\bf 95} 115410 (2017). DOI:10.1103/PhysRevB.95.115410, arXiv:1508.03114. 
 
\bibitem{Peeters2013} 
M. Neek-Amal, L. Covaci, K. Shakouri, F. M. Peeters, \textit{Electronic structure of a hexagonal graphene flake subjected to triaxial stress}, Phys. Rev. B {\bf 88} 115428 (2013). DOI:10.1103/PhysRevB.88.115428, arXiv:1404.4966. 
 
\bibitem{Peeters2013_2} 
D. Moldovan, M. R. Masir, F. M. Peeters, \textit{Electronic states in a graphene flake strained by a Gaussian bump}, Phys. Rev. B {\bf 88} 035446 (2013). DOI:10.1103/PhysRevB.88.035446, arXiv:1307.5190. 
 
\bibitem{CastroNeto2014}  
Z. Qi, A. L. Kitt, H. S. Park, V. M. Pereira, D. K. Campbell, A. H. Castro Neto, \textit{Pseudomagnetic fields in graphene nanobubbles of constrained geometry: A molecular dynamics study}, Phys. Rev. B {\bf 90} 125419 (2014). DOI:10.1103/PhysRevB.90.125419, arXiv:1406.1092. 
 
\bibitem{Zhu2014} 
S. Zhu, Y. Huang, N. N. Klimov, D. B. Newell, N. B. Zhitenev, J. A. Stroscio, S. D. Solares, T. Li, \textit{Pseudomagnetic fields in a locally strained graphene drumhead}, Phys. Rev. B {\bf 90} 075426 (2014). DOI:10.1103/PhysRevB.90.075426, arXiv:1505.02805. 
 
 
 
 
 
 
 
\bibitem{Cvetic:2012vg} 
G. W. Gibbons, M. Cvetic, ``Graphene and the Zermelo optical metric of the
BTZ black hole'', Ann. Phys. {\bf 327},  2617-2626  (2012). DOI:10.1016/j.aop.2012.05.013, arXiv:1202.2938. 
 
\bibitem{Gibbons:2011rh}
  G.~W.~Gibbons and M.~Vyska,
``The Application of Weierstrass elliptic functions to Schwarzschild Null Geodesics'',   Class.\ Quant.\ Grav.\  {\bf 29} (2012) 065016. DOI:10.1088/0264-9381/29/6/065016, arXiv:1110.6508. 
 
\bibitem{LevyLeblond:1967zz}
  J.~M.~Levy-Leblond, ``Nonrelativistic particles and wave equations'', Commun.\ Math.\ Phys.\  {\bf 6} (1967) 286.   DOI:10.1007/BF01646020.  
  

\bibitem{Horvathy:1993fd}
 P.~A.~Horvathy,    ``Non-Relativistic Conformal and Supersymmetries'', Int.\ J.\ Mod.\ Phys.\ A {\bf 3} (1993) 339.  arXiv:0807.0513. 

\bibitem{Horvathy1996} 
C. Duval, P. A. Horv\'athy and L. Palla,  ``Spinors in non-relativistic Chern-Simons electrodynamics'', Ann. Phys. {\bf 249}, 265 (1996). DOI:10.1006/aphy.1996.0071, arXiv:hep-th/9510114. 
   
\bibitem{Cariglia2012} 
M. Cariglia, ``Hidden symmetries of Eisenhart-Duval lift metrics and the Dirac equation with flux'', Phys. Rev. D {\bf 86} (2012) 084050. DOI:10.1103/PhysRevD.86.084050,arXiv:1206.0022. 
  
\bibitem{Gary1982} 
G. W. Gibbons, ``The multiplet structure of solitons in the O(2) supergravity theory'', London 1981, Proceedings, Quantum Structure Of Space and Time, 31-321 Conference: C81-08-03.3. 
 
 \bibitem{Gibbons:1981ux}
  G.~W.~Gibbons, ``Supersymmetric Soliton States and Central Charges in Extended Supergravity Theories'', Lect.\ Notes Phys.\  {\bf 160} (1982) 145.
 
 \bibitem{Gibbons:2008rs}
G.~W.~Gibbons and M.~Rogatko, ``The Decay of Dirac Hair around a Dilaton Black Hole'',
  Phys.\ Rev.\ D {\bf 77} (2008) 044034.  DOI:10.1103/PhysRevD.77.044034, arXiv:0801.3130. 
   
\bibitem{GonzalezHerrero2010}
J. Gonz\'alez, J. Herrero, ``Graphene wormholes: A condensed matter illustration of Dirac fermions in curved space'', Nucl. Phys. B. {\bf 825} (2010) 426. DOI:10.1016/j.nuclphysb.2009.09.028, arXiv:0909.3057. 
 


 
\bibitem{Atanasov2011}
V. Atanasov, A. Saxena, ``Electronic properties of corrugated graphene: the Heisenberg principle and wormhole geometry in the solid state'', J. Phys.: Cond. Matt. {\bf 23} (2011) 175301. DOI:10.1088/0953-8984/23/17/175301, arXiv:1101.5243. 
 
\bibitem{PincakSmotlacha2013} 
R. Pincak, J. Smotlacha, ``Analogies in electronic properties of graphene wormhole and perturbed nanocylinder'', Eur. Phys. J. B {\bf 86} (2013) 480. DOI:10.1140/epjb/e2013-40594-0, arXiv:1511.02748.  
  
\bibitem{Capozziello2017} 
A. Sepehri, R. Pincak, K. Bamba, S. Capozziello, E. N. Saridakis, ``Current density and conductivity through modified gravity in the graphene with defects'', Int. J. Mod. Phys. D {\bf 26} (2017) 1750094. DOI:10.1142/S0218271817500948, arXiv:1607.01499. 
  
\bibitem{STM2012} 
E. Y. Andrei, G. Li, X. Du, ``Electronic properties of graphene: a perspective from scanning tunneling microscopy and magnetotransport'', Rep. Progr. Phys. {\bf 75} (2012) 056501. DOI:10.1088/0034-4885/75/5/056501, arXiv:1204.4532.  
 


\end{thebibliography}
\end{document}